\documentclass{article}
\usepackage{spconf,amsmath,graphicx,bm, amssymb}
\usepackage{enumitem}

\usepackage{hyperref}
\usepackage{booktabs, multirow}
\usepackage{bibspacing}

\usepackage[numbers, sort&compress]{natbib}
\usepackage[linesnumbered,ruled,vlined]{algorithm2e}

\def\dd{{\mathrm d}}

\title{VoiceFlow: Efficient Text-to-Speech with Rectified Flow Matching}
\name{Yiwei Guo, Chenpeng Du, Ziyang Ma, Xie Chen, Kai Yu\sthanks{Corresponding author}}
\address{X-LANCE Lab, Department of Computer Science and Engineering\\
Shanghai Jiao Tong University, Shanghai, China\\
MoE Key Lab of Artificial Intelligence, AI Institute
\\ {\texttt{\{yiwei.guo, duchenpeng, zym.22, chenxie95, kai.yu\}@sjtu.edu.cn}}}
\begin{document}
\ninept
\maketitle
\begin{abstract}
Although diffusion models in text-to-speech have become a popular choice due to their strong generative ability, the intrinsic complexity of sampling from diffusion models harms their efficiency.
Alternatively, we propose VoiceFlow, an acoustic model that utilizes a rectified flow matching algorithm to achieve high synthesis quality with a limited number of sampling steps.
VoiceFlow formulates the process of generating mel-spectrograms into an ordinary differential equation conditional on text inputs, whose vector field is then estimated.
The rectified flow technique then effectively straightens its sampling trajectory for efficient synthesis.
Subjective and objective evaluations on both single and multi-speaker corpora showed the superior synthesis quality of VoiceFlow compared to the diffusion counterpart.
Ablation studies further verified the validity of the rectified flow technique in VoiceFlow.
\end{abstract}
\vspace{-.1in}
\section{Introduction}

\label{sec:intro}

Modern text-to-speech (TTS) has witnessed tremendous progress by adopting different types of advanced generative algorithms, such as TTS models with GANs~\cite{binkowski2019high,kim2021conditional}, normalizing flows~\cite{kim2020glow,valle2021flowtron,kim2021conditional}, self-supervised features~\cite{du2022vqtts,du2023speaker} or denoising diffusion models~\cite{popov2021grad,liu2022diffsinger,liu2023diffvoice,du2023unicats}.
Among them, diffusion-based TTS models recently received growing attention because of their high synthesis quality, such as GradTTS~\cite{popov2021grad} and DiffVoice~\cite{liu2023diffvoice}.
They also show versatile functionalities such as conditional generation~\cite{kim2022guided,guo2023emodiff}, speech editing~\cite{tae2021editts,liu2023diffvoice, du2023unicats} and speaker adaptation~\cite{liu2023diffvoice, du2023unicats}.
By estimating the score function $\nabla \log p_t(\bm x)$ of a stochastic differential equation (SDE), diffusion models are stable to train~\cite{dhariwal2021diffusion}.
They generate realistic samples by numerically solving the reverse-time SDE or the associated probability-flow ordinary differential equation (ODE).

However, a major drawback of diffusion models lies in their efficiency.
Regardless of SDE or ODE sampling methods, diffusion models typically require numerous steps to generate a satisfying sample,
causing a large latency in inference.
Some efforts have been made to mitigate this issue and improve the speed-quality tradeoff in diffusion-based TTS models, usually by extra mathematical tools or knowledge distillation.
Fast GradTTS\cite{vovk2022fast} adopts maximum likelihood SDE solver~\cite{popov2022diffusionbased}, progressive distillation~\cite{salimans2022progressive} and denoising diffusion GAN~\cite{xiao2022tackling} to accelerate diffusion sampling.
FastDiff~\cite{huang2022fastdiff} optimizes diffusion noise schedules inspired by BDDM~\cite{lam2022bddm}.
ProDiff~\cite{huang2022prodiff} also uses a progressive distillation technique to halve the sampling steps from DDIM~\cite{song2021denoising} teacher iteratively.
LightGrad~\cite{chen2023lightgrad} adopts DPM-Solver~\cite{lu2022dpm} to explicitly derive a solution of probability-flow ODE.
A concurrent work, CoMoSpeech~\cite{ye2023comospeech}, integrates the consistency model~\cite{song2023consistency} as a special type of diffusion distillation.
These models successfully decrease the necessary number of sampling steps in diffusion models to some extent.
However, due to the intricate nature of the diffusion process, the speed-quality tradeoff still exists and is hard to overcome.

Despite denoising diffusion, another branch in the family of differential-equation-based generative models began to arise recently, namely the flow matching generative models~\cite{lipman2022flow,liu2022flow,tong2023improving}.
While diffusion models learn the score function of a specific SDE, flow matching aims to model the vector field implied by an arbitrary ODE directly.
A neural network is used for approximating the vector field, and the ODE can also be numerically solved to obtain data samples.
The design of such ODE and vector field often considers linearizing the sampling trajectory and minimizing the transport cost~\cite{tong2023improving}.
As a result, flow matching models have simpler formulations and fewer constraints but better quality.
VoiceBox~\cite{le2023voicebox} shows the potential of flow matching in fitting large-scale speech data, and LinDiff~\cite{liu2023boosting} shares a similar concept in the study of vocoders.
More importantly, the rectified flow~\cite{liu2022flow} technique in flow matching models further straightens the ODE trajectory in a concise way.
By training a flow matching model again but with its own generated samples, the sampling trajectory of rectified flow theoretically approaches a straightforward line, which improves the efficiency of sampling.
In essence, rectified flow matching achieves good sample quality even with a very limited number of sampling steps.
As a side note, its ODE nature also makes flow matching extensible for knowledge distillation similar in previous diffusion-based works~\cite{liu2022flow}.

Inspired by these, we propose to utilize rectified flow matching in the TTS acoustic model for the first time in literature.
We construct an ODE to flow between noise distribution and mel-spectrogram while conditioning it with phones and duration.
An estimator learns to model the underlying vector field.
Then, a flow rectification process is applied, where we generate samples from the trained flow matching model to train itself again.
In this way, our model is able to generate decent mel-spectrograms with much fewer steps.
We name our model VoiceFlow.
To fully investigate its ability, we experiment both on the single-speaker benchmark LJSpeech and the larger multi-speaker dataset LibriTTS.
The results show that VoiceFlow outperforms the diffusion baseline in a sufficient number of sampling steps.
In a highly limited budget such as two steps, VoiceFlow still maintains a similar performance while the diffusion model cannot generate reasonable speech.
Therefore, VoiceFlow achieves better efficiency and speed-quality tradeoff while sampling.
The code and audio samples are available online\footnote{\url{https://cantabile-kwok.github.io/VoiceFlow}}.


\section{Flow Matching and Rectified Flow}
\label{sec:flow-matching}
\subsection{Flow Matching Generative Models}
Denote the data distribution as $p_1(\bm x_1)$ and some tractable prior distribution as $p_0(\bm x_0)$.
Most generative models work by finding a way to map samples $\bm x_0\sim p_0(\bm x_0)$ to data $\bm x_1$.
Particularly, diffusion models manually construct a special SDE, and then estimate the score function of the probability path $p_t(\bm x_t)$ yielded by it.
Sampling is tackled by solving either the reverse-time SDE or probability-flow ODE alongside this probability path.
Flow matching generative models, on the other hand, model the probability path $p_t(\bm x_t)$ directly~\cite{lipman2022flow}.
Consider an arbitrary ODE
\begin{equation}
    \label{eq:ODE}
    {\dd \bm x_t} =\bm v_t(\bm x_t){\dd t}
\end{equation}
with $\bm v_t(\cdot)$ named the vector field and $t\in[0,1]$.
This ODE is associated with a probability path $p_t(\bm x_t)$ by the continuity equation $\frac{\dd }{\dd t} \log p_t(\bm x) + \operatorname{div}(p_t(\bm x)\bm v_t(\bm x)) = 0$.
It is sufficient to generate realistic data if a neural network can accurately estimate the vector field $\bm v_t(\cdot)$, since the ODE in Eq.\eqref{eq:ODE} can be solved numerically then.

However, the design of the vector field needs to be instantiated before practically applied.
\cite{lipman2022flow} proposes the method of constructing a conditional probability path with a data sample $\bm x_1$.
Suppose this probability path is $p_t(\bm x\mid \bm x_1)$, with boundary condition $p_{t=0}(\bm x\mid \bm x_1) = p_0(\bm x)$ and $p_{t=1}(\bm x\mid\bm x_1) = \mathcal N(\bm x\mid \bm x_1, \sigma^2\bm I)$ for sufficiently small $\sigma$.
By the continuity equation, there is an associated vector field $\bm v_t(\bm x\mid \bm x_1)$.
It is proven that estimating the conditional vector field by neural network $\bm u_\theta$ is equivalent, in the sense of expectation, to estimating the unconditional vector field, i.e.
\begin{align}
    &\min_{\theta}\mathbb E_{t,p_t(\bm x)}\|\bm u_\theta(\bm x,t) - \bm v_t(\bm x)\|^2\\
    \equiv &\min_{\theta}\mathbb E_{t,p_1(\bm x_1),p_t(\bm x\mid \bm x_1)}\|\bm u_\theta(\bm x,t) - \bm v_t(\bm x\mid \bm x_1)\|^2.
    \label{eq:fm-target}
\end{align}
Then, by designing a simple conditional probability path $p_t(\bm x\mid \bm x_1)$ and the corresponding $\bm v_t(\bm x\mid \bm x_1)$, one can easily draw samples from $p_t(\bm x\mid \bm x_1)$ and minimize Eq.\eqref{eq:fm-target}.
For example, \cite{lipman2022flow} uses the Gaussian path $p_t(\bm x\mid \bm x_1)=\mathcal N(\bm x\mid \bm \mu_t(\bm x_1),\sigma_t(\bm x_1)^2\bm I)$ and linear vector field $\bm v_t(\bm x\mid \bm x_1)=\frac{\sigma'_t(\bm x_1)}{\sigma_t(\bm x_1)}(\bm x-\bm \mu_t(\bm x_1))+\bm \mu_t'(\bm x_1)$.

Meanwhile, this conditioning technique can be further generalized, i.e. any condition $z$ for $p_t(\bm x\mid z)$ can lead to the same form of optimization target like Eq.\eqref{eq:fm-target}.
Thus, \cite{tong2023improving} proposes to additionally condition on a noise sample $\bm x_0$ to form a probability path $p_t(\bm x\mid \bm x_0, \bm x_1)=\mathcal N(\bm x\mid t\bm x_1 + (1-t)\bm x_0, \sigma^2 \bm I)$.
The conditional vector field therefore becomes $\bm v_t(\bm x\mid \bm x_0, \bm x_1)=\bm x_1-\bm x_0$, which is a constant straight line towards $\bm x_1$.
In this formulation, training the generative model only requires the following steps:
\begin{enumerate}[leftmargin=*, noitemsep, topsep=5pt]
    \item Sample $\bm x_1$ from data and $\bm x_0$ from any noise distribution $p_0(\bm x_0)$;
    \item Sample a time $t\in[0,1]$ and then $\bm x_t \sim \mathcal N(t\bm x_1 + (1-t)\bm x_0, \sigma^2 \bm I)$;
    \item Apply gradient descent on loss $\|\bm u_\theta(\bm x,t) - (\bm x_1 - \bm x_0)\|^2$.
\end{enumerate}
This is often referred to as the ``conditional flow matching" algorithm, which is proven to outperform diffusion-based models with deep correlation to the optimal transport theory~\cite{tong2023improving}. 

\subsection{Improved Sampling Efficiency with Rectified Flow}
The notion of rectified flow is proposed in \cite{liu2022flow}.
It is a simple but mathematically solid approach to improve the sampling efficiency of flow matching models.
The flow matching model here has the same formulation as that of \cite{tong2023improving}, which is conditioned on both $\bm x_1$ and $\bm x_0$.
Suppose a flow matching model is trained to generate data $\bm {\hat x}_1$ from noise $\bm x_0$ by the ODE in Eq.\eqref{eq:ODE}. In other words, $\bm x_0$ and $\bm {\hat x}_1$ are a pair of the starting and ending points of the ODE trajectory.
Then, this flow matching model is trained again, but conditions $\bm v_t(\bm x\mid \bm x_0,\bm x_1)$ and $p_t(\bm x\mid \bm x_0,\bm x_1)$ on the given pair $(\bm x_0, \bm {\hat x}_1)$ instead of independently sampling $\bm x_0, \bm x_1$.
This flow rectification step can be iterated multiple times, denoted by the recursion $\left(\bm z_0^{k+1}, \bm z_1^{k+1}\right) = \operatorname{FM}\left(\bm z_0^{k}, \bm z_1^k\right)$, with $\operatorname{FM}$ the flow matching model and $(\bm z_0^0, \bm z_1^0)=(\bm x_0, \bm x_1)$ the independently drawn noise and data samples.

Intuitively, rectified flow ``rewires" the sampling trajectory of flow matching models to become more straight.
Because the ODE trajectories cannot intersect when being solved, most likely the trajectory cannot be as straight as the conditional vector field in training. 
However, by training the flow matching model again on the endpoints of the same trajectory, the model learns to find a shorter path to connect these noise and data.
This straightening tendency is theoretically guaranteed in \cite{liu2022flow}.
By rectifying the trajectories, flow matching models will be able to sample data more efficiently with fewer steps of ODE simulation.

\section{VoiceFlow}
\begin{figure}
    \centering
    \includegraphics[width=0.99\linewidth]{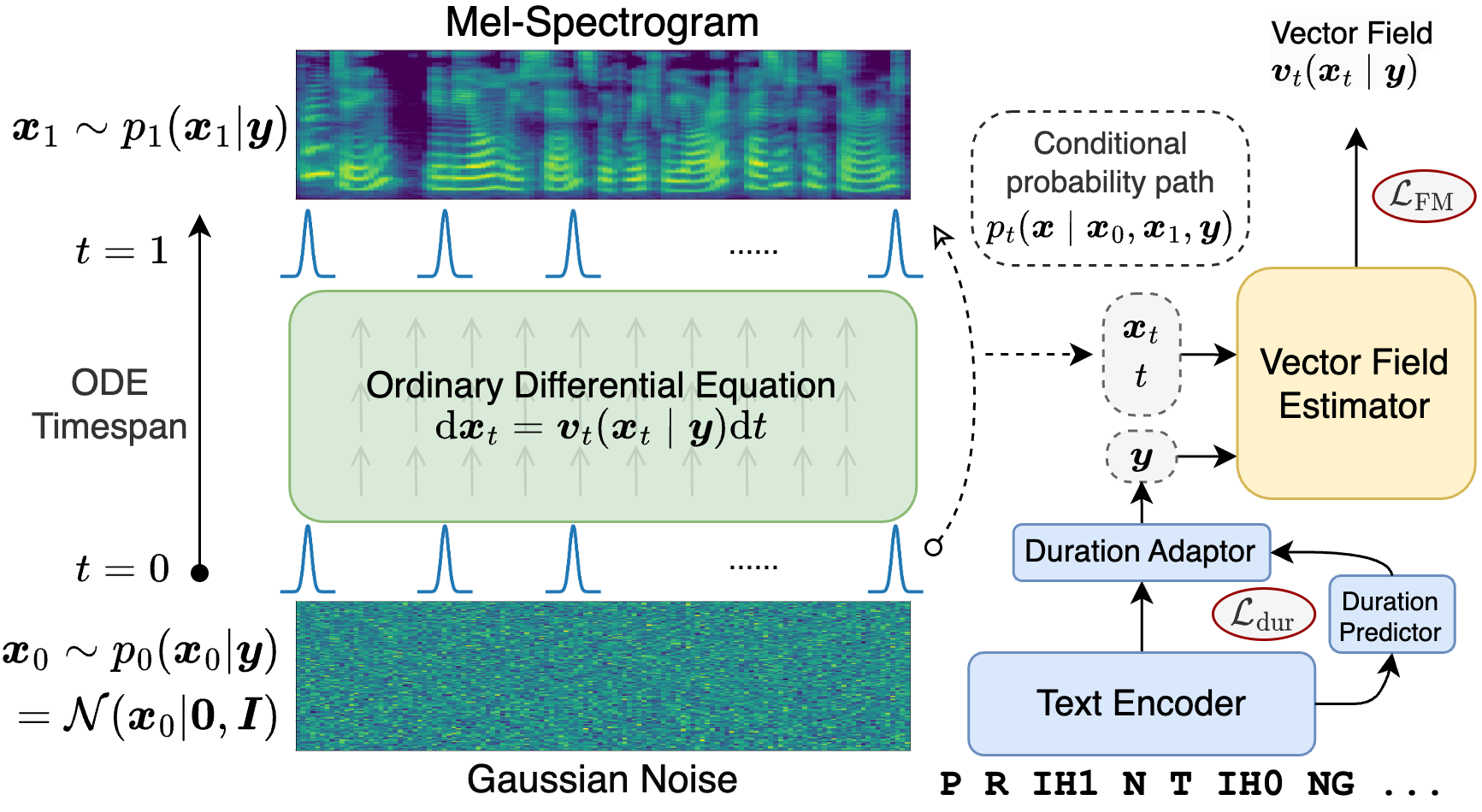}
    \vspace{-0.1in}
    \caption{Working diagram of the VoiceFlow model}
    \label{fig:model}
\end{figure}
\label{sec:model}
\subsection{Flow Matching-Based Acoustic Model}
To utilize flow matching models in TTS, we cast it as a non-autoregressive conditional generation problem with mel-spectrogram $\bm x_1\in \mathbb R^d$ as the target data and noise $\bm x_0\in\mathbb R^d$ from standard Gaussian distribution $\mathcal N(\bm 0, \bm I)$.
We consider using an explicit duration learning module from forced alignments like in \cite{liu2022diffsinger}.
Denote the duplicated latent phone representation as $\bm y$, where each phone's latent embedding is repeated according to its duration.
Then, $\bm y$ is regarded as the condition of the generation process.
Specifically, suppose $\bm v_t(\bm x_t\mid\bm y)\in \mathbb R^d$ is the underlying vector field for the ODE $\dd \bm x_t = \bm v_t(\bm x_t\mid \bm y)\dd t$.
Suppose this ODE connects the noise distribution $p_0(\bm x_0\mid \bm y)=\mathcal N(\bm 0, \bm I)$ with mel distribution given text $p_1(\bm x_1\mid \bm y) = p_{\text{mel}}(\bm x_1\mid \bm y)$.
Our goal is to accurately estimate the vector field $\bm v_t$ given condition $\bm y$, as we can then generate a mel-spectrogram by solving this ODE from $t=0$ to $t=1$.

Inspired by \cite{liu2022flow,tong2023improving}, we opt to use both a noise sample $\bm x_0$ and a data sample $\bm x_1$ to construct conditional probability paths as
\begin{equation}\label{eq:cond-path-in-voiceflow}
    p_t(\bm x\mid \bm x_0,\bm x_1, \bm y)=\mathcal N(\bm x\mid t\bm x_1+(1-t)\bm x_0, \sigma^2\bm I)
\end{equation}
where $\sigma$ is a sufficiently small constant.
In this formulation, the endpoints of these paths are $\mathcal N(\bm x_0,\sigma^2\bm I)$ for $t=0$ and $\mathcal N(\bm x_1,\sigma^2\bm I)$ for $t=1$ respectively.
These paths also determine a probability path $p_t(\bm x\mid \bm y)$ marginal w.r.t $\bm x_0,\bm x_1$, whose boundaries approximate the noise distribution $p_0(\bm x_0\mid \bm y)$ and mel distribution $p_1(\bm x_1\mid \bm y)$.
Intuitively, Eq.\eqref{eq:cond-path-in-voiceflow} specifies a family of Gaussians moving in a linear path.
The related vector field can be simply $\bm v_t(\bm x\mid \bm x_0, \bm x_1, \bm y)=\bm x_1-\bm x_0$, also a constant linear line.

Then, we use a neural network $\bm u_\theta$ to estimate the vector field.
Similar to Eq.\eqref{eq:fm-target}, the objective here is
\begin{equation}\label{eq:objective-voiceflow}
    \min_\theta \mathbb E_{t,p_1(\bm x_1\mid \bm y), p_0(\bm x_0\mid \bm y),p_t(\bm x_t\mid \bm x_0, \bm x_1, \bm y)}\|\bm u_\theta(\bm x_t,\bm y, t)-(\bm x_1-\bm x_0)\|^2
\end{equation}
The corresponding flow matching loss is denoted by $\mathcal L_{\text{FM}}$.
The total loss function to train VoiceFlow will be $\mathcal L=\mathcal L_{\text{FM}} + \mathcal L_{\text{dur}}$, where $\mathcal L_{\text{dur}}$ is the mean squared loss for duration predictor.
So, the whole acoustic model of VoiceFlow consists of the text encoder, duration predictor, duration adaptor and vector field estimator, as is shown in Fig. \ref{fig:model}.
The text encoder transforms the input phones into a latent space, upon which the duration per phone is predicted and fed to the duration adaptor.
The repeated frame-level sequence $\bm y$ is then fed to the vector field estimator as a condition.
The other two inputs to the vector field estimator are the sampled time $t$ and the sampled $\bm x_t$ from the conditional probability path in Eq.\eqref{eq:cond-path-in-voiceflow}.
We adopt the same U-Net architecture in the vector field estimator as in GradTTS\footnote{Two down and up-samples with 2D convolution as residual blocks}.
The condition $\bm y$ is concatenated with $\bm x_t$ before entering the estimator, and the time $t$ is passed through some fully connected layers before being added to the hidden variable in residual blocks each time.

In multi-speaker scenarios, the condition will become both the text $\bm y$ and some speaker representation $\bm s$.
But for simplicity, we will still use the notation of $\bm y$ as the condition in the following sections.

\subsection{Sampling and Flow Rectification Step}
By Eq.\eqref{eq:objective-voiceflow}, the vector field estimator $\bm u_\theta$ is able to approximate $\bm v_t$ in the expectation sense.
Then, the ODE $\dd \bm x_t = \bm u_\theta(\bm x_t, \bm y, t)\dd t$ can be discretized for sampling a synthetic mel-spectrogram $\bm x_1$ given text $\bm y$.
Off-the-shelf ODE solvers like Euler, Runge-Kutta, Dormand-Prince method, etc. can be directly applied for sampling. 
In the example of the Euler method with $N$ steps, each sampling step is
\begin{equation}\label{eq:solver}
    \bm {\hat x}_{\frac{k+1}{N}} = \bm {\hat x}_{\frac kN} + \frac1N{\bm u_\theta\left(\bm {\hat x}_{\frac kN} , \bm y, \frac kN\right)}, k=0,1,...,N-1
\end{equation}
with $\bm {\hat x}_0\sim p_0(\bm x_0\mid \bm y)$ being the initial point and $\bm {\hat x}_1$ being the generated sample.
Regardless of the discretization method, the solvers will produce a sequence of samples $\{\bm {\hat x}_{k/N}\}$ along the ODE trajectory, which gradually approximates a realistic spectrogram.

Then we apply the rectified flow technique to further straighten the ODE trajectory.
For every utterance in the training set, we draw a noise sample $\bm x'_0$ and run the ODE solver to obtain $\bm {\hat x}_1$ given text $\bm y$.
The sample pair $(\bm x'_0, \bm {\hat x}_1)$ is then fed to the VoiceFlow again for rectifying the vector field estimator.
In this flow rectification step, the new training criterion will be
\begin{equation}\label{eq:objective-voiceflow-reflow}
    \min_\theta \mathbb E_{t,p(\bm x'_0, \bm {\hat x}_1\mid \bm y),p_t(\bm x_t\mid \bm x'_0,\bm {\hat x}_1,\bm y)} \|\bm u_\theta(\bm x_t, \bm y , t) - (\bm {\hat x}_1-\bm x'_0)\|^2
\end{equation}
where the only difference with Eq.\eqref{eq:objective-voiceflow} is paired $(\bm x'_0, \bm {\hat x}_1)$ are used instead of independently sampled.
In Eq.\eqref{eq:objective-voiceflow-reflow}, every spectrogram sample $\bm {\hat x}_1$ is associated with a noise sample in the same trajectory.
In this way, the vector field estimator is asked to find a more straightforward sampling trajectory connecting $(\bm x'_0, \bm {\hat x}_1)$, which improves the sampling efficiency to a large extent.
Note that we provide the model with the ground truth duration sequence while generating data for rectified flow.
This ensures that the model is fed with more natural speech, reducing the risk that inaccurate duration prediction degrades the model performance.

Algorithm \ref{algo} summarizes the whole process of training VoiceFlow,
including flow rectification.
\begin{algorithm}
\label{algo}
\SetInd{0em}{1em}
\KwIn{Paired text, duration and mel-spectrogram $\bm x_1$ with optional speaker information}
\KwResult{Trained VoiceFlow model, which contains vector field estimator $\bm u_\theta(\bm x,\bm y,t)$}
\SetKwFunction{TrainStep}{TrainStep}
\SetKwProg{Fn}{Function}{:}{}
\Fn{\TrainStep{$\bm u_\theta,\bm x_0,\bm x_1$}}{
Compute $\bm y, \mathcal L_{\text{dur}}$ using text and duration\;
    Sample $t\sim \operatorname{Uniform}[0,1]$\;
    Sample $\bm x_t\sim \mathcal N(t\bm x_1+(1-t)\bm x_0, \sigma^2\bm I)$\;
    $\mathcal L_{\text{FM}}\leftarrow\|\bm u_\theta(\bm x_t,\bm y,t)-(\bm x_1-\bm x_0)\|^2$\;
     Gradient descent on $\mathcal L_{\text{FM}}+\mathcal L_{\text{dur}}$\;
  }
 \While{perform flow matching}{
  Take batch and sample $\bm x_0$ from $\mathcal N(\bm 0,\bm I)$\;
  \TrainStep{$\bm u_\theta,\bm x_0, \bm x_1$}\;
 }
 \For{every utterance in training set}{
     Sample a $\bm x'_0$ and solve the ODE by Eq.\eqref{eq:solver} to obtain $\bm {\hat x}_1$, using trained $\bm u_\theta$ and ground truth duration\;
 }
 \While{perform flow rectification}{
 Take batch with associated $\bm x'_0$\;
 \TrainStep{$\bm u_\theta, \bm x'_0,\bm {\hat x_1}$}\;
    }
 \caption{Training VoiceFlow with flow rectification}
\end{algorithm}

\vspace{-0.2in}
\section{Experiments and Results}
\label{sec:exp}
\subsection{Experimental Setup}
We evaluated VoiceFlow both on the single-speaker and multi-speaker benchmarks, so as to obtain a comprehensive observation of the proposed TTS system.
For single-speaker evaluations, we used the LJSpeech~\cite{ito2017lj} dataset, which contains approximately 24 hours of high-quality female voice recordings.
For multi-speaker experiments, we included all the training partitions of the LibriTTS~\cite{libritts} dataset, which amounted to 585 hours and over 2300 speakers.
We downsampled all the training data to 16kHz for simplicity.
Mel-spectrogram and forced alignments were extracted with 12.5ms frame shift and 50ms frame length on each corpus by Kaldi~\cite{povey2011kaldi}.

We compared VoiceFlow with the diffusion-based acoustic model GradTTS.
To only focus on the algorithmic differences,
we used the official implementation of GradTTS\footnote{\url{https://github.com/huawei-noah/Speech-Backbones/tree/main/Grad-TTS}} and trained it with the same data configurations.
We also used ground truth duration instead of the monotonic alignment search algorithm in GradTTS to mitigate the impact of different durations.
Notably, we used exactly the same model architecture to build VoiceFlow, so the two compared models have nearly identical inference costs when the ODEs are both solved using Euler method with the same number of steps.

As the acoustic models generate mel-spectrograms as targets, HifiGAN~\cite{kong2020hifi} was adopted as the vocoder and trained separately on the two datasets.

\vspace{-.1in}
\subsection{Subjective Evaluations}
We first evaluated the system performance of VoiceFlow compared to GradTTS using subjective listening tests.
In this test, listeners were asked to rate the mean opinion score (MOS) of the provided speech clips based on the audio quality and naturalness.
For both the acoustic models, we used 2, 10 and 100 steps representing low, medium and large number of sampling steps for synthesis.
The results are presented in Table \ref{tab:naturalness}, where ``GT (voc.)" means the vocoded ground truth speech.
It can be seen that in both the two datasets and three sampling scenarios, VoiceFlow achieves a consistently higher MOS score than GradTTS.
Also, when the sampling steps are decreased, the performance of GradTTS drops significantly while VoiceFlow does not suffer from such huge degeneration.
Specifically, in 2-step sampling situations, samples from GradTTS become heavily degraded, but that from VoiceFlow remains to be satisfying.
Note that the 10-step GradTTS was already reported to be competitive against other baselines~\cite{popov2021grad}.
In LibriTTS, the corpus with large speaker and environment variability, the difference of compared systems becomes more obvious.
This suggests the stronger potential of flow-matching-based models in fitting complex speech data.

\begin{table}[]
\centering
\caption{MOS evaluation in low, medium and large sampling steps}
\label{tab:naturalness}
\begin{tabular}{@{}lcll@{}}
\toprule
\multicolumn{1}{c}{\textbf{Model}} & \textbf{Sampling Steps} & \multicolumn{1}{c}{\textbf{LJSpeech}} & \multicolumn{1}{c}{\textbf{LibriTTS}} \\ \midrule
GradTTS & 2  & 2.98$\pm$0.06 & 2.52$\pm$0.12 \\
VoiceFlow & \footnotesize{($\approx$3605frames/s)} &  3.92$\pm$0.07&  3.81$\pm$0.07\\ \cmidrule(r){1-2}
GradTTS & 10 &  3.97$\pm$0.07&3.43$\pm$0.09  \\
VoiceFlow &  \footnotesize{($\approx$985frames/s)} &4.10$\pm$0.06  &  3.84$\pm$0.07\\ \cmidrule(r){1-2}
GradTTS & 100 & 4.03$\pm$0.09 &  3.45$\pm$0.12\\
VoiceFlow & \footnotesize{($\approx$102frames/s)} & \textbf{4.17$\pm$0.07} &  \textbf{3.85$\pm$0.12}\\ \midrule
GT (voc.) & - &  4.52$\pm$0.07&  4.42$\pm$0.06\\ \bottomrule
\end{tabular}
\end{table}

\vspace{-0.05in}
\subsection{Objective Evaluations}
\begin{figure}
    \centering
    \includegraphics[width=0.99\linewidth]{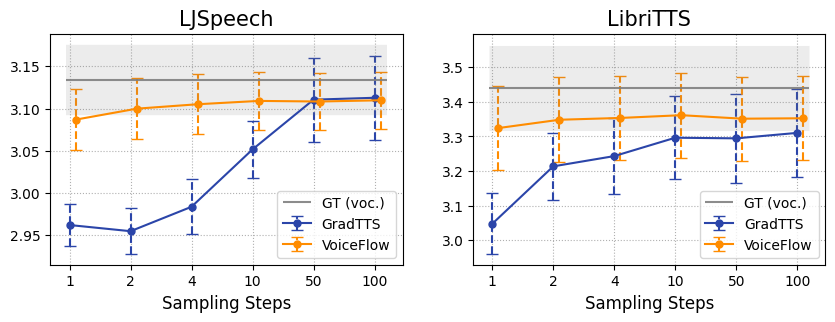}
    \vspace{-0.1in}
    \caption{MOSnet evaluations in multiple choices of sampling steps}
\vspace{-0.2in}
    \label{fig:MOSnet}
\end{figure}

\begin{figure}
    \centering
    \vspace{-0.1in}
    \includegraphics[width=0.99\linewidth]{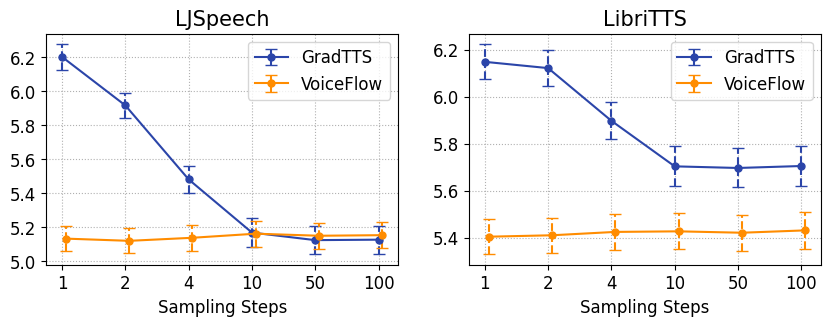}
    \vspace{-0.1in}
    \caption{MCD evaluations in multiple choices of sampling steps}
    \label{fig:MCD}
\end{figure}
\vspace{-0.05in}
We also objectively evaluated the performance of VoiceFlow.
Two metrics were included for comparison: MOSnet~\cite{mosnet} and mel-cepstral distortion (MCD).
MOSnet is a neural network designed to fit human perception of speech signals, and we found it correctly reflects speech quality to a reasonable extent.
We use the officially trained MOSnet model to evaluate synthetic speech on more choices of sampling steps.
The results are plotted in Fig. \ref{fig:MOSnet}, where the shadowed region stands for the mean and 95\% confidence interval of MOSnet score on ground truth speech.
It can be seen to mainly conform with the MOS results, as the change of VoiceFlow's scores among different sampling steps is much lower than that of GradTTS.
MCD is another objective tool to measure the distortion of the cepstrum against ground truth.
The cepstrum order here is set to be 13.
Similarly, the MCD values on different numbers of sampling steps are shown in Fig. \ref{fig:MCD}, also verifying the better speed-quality tradeoff of VoiceFlow compared to the diffusion counterpart.

\vspace{-0.1in}
\subsection{Ablation Study}
\begin{table}[]
\centering
\caption{CMOS evaluation in 2 sampling steps}
\label{tab:cmos}
\begin{tabular}{@{}lll@{}}
\toprule
\multicolumn{1}{c}{\textbf{Model}} & \multicolumn{1}{c}{\textbf{LJSpeech}} & \multicolumn{1}{c}{\textbf{LibriTTS}} \\ \midrule
VoiceFlow & - & - \\
\multicolumn{1}{r}{-ReFlow} & -0.78$\pm$0.13 & -1.21$\pm$0.19 \\ \bottomrule
\end{tabular}
\end{table}

We also conducted an ablation study to verify the effectiveness of the rectified flow technique in VoiceFlow.
A comparative MOS (CMOS) test was performed where raters were asked to rate the score of a given sentence compared to the reference, ranging from -3 to 3.
Table \ref{tab:cmos} shows the results with 2 sampling steps, where ``-ReFlow" means VoiceFlow without rectified flow.
It is noticeable that rectified flow makes a remarkable effect in such limited sampling steps, and LibriTTS exhibits an even more significant difference than LJSpeech.

\begin{figure}
    \centering
    \includegraphics[width=0.99\linewidth]{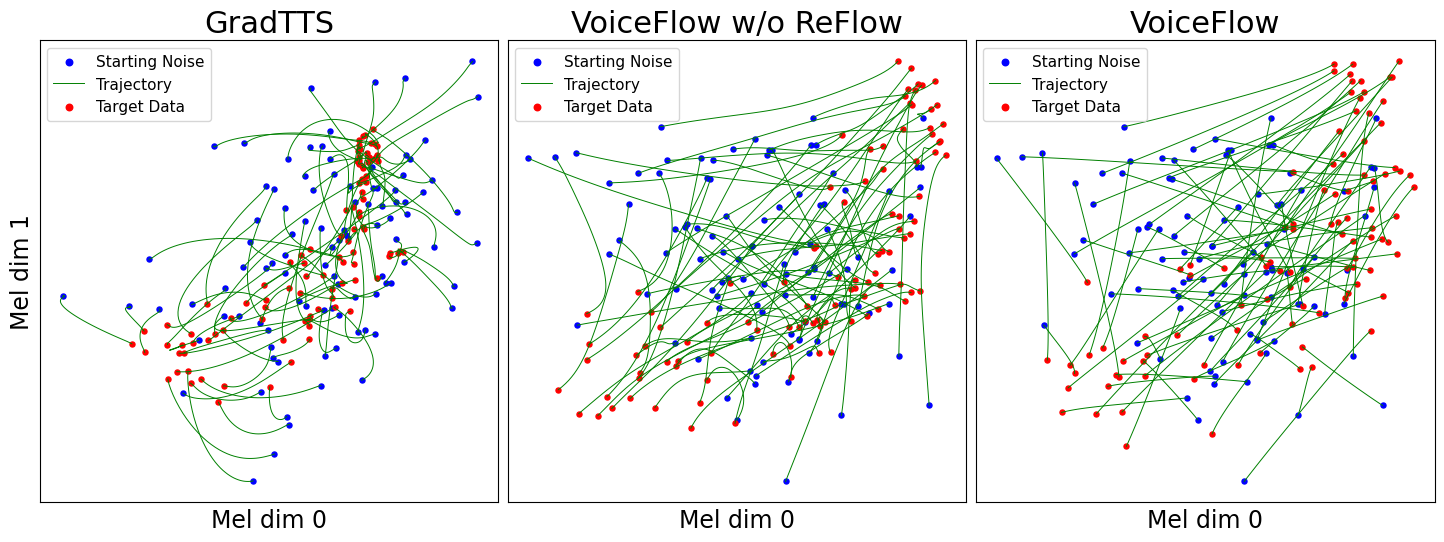}
    \vspace{-0.15in}
    \caption{Visualization of sampling trajectories}
    \vspace{-0.2in}
    \label{fig:traj}
\end{figure}

To provide an intuition on the impact of rectified flow, we visualized some sampling trajectories of VoiceFlow both with and without rectified flow on two out of the 80 mel dimensions in Figure \ref{fig:traj}.
The trajectory of GradTTS is also shown here.
Then, the visual contrast between the straight and curving trajectories leaves no doubt on the efficacy of using rectified flow in TTS models.

\vspace{-0.1in}
\section{Conclusion}
\label{sec:conclusion}
In this study, we proposed VoiceFlow, a TTS acoustic model based on rectified flow matching, that achieves better efficiency and speed-quality tradeoff than diffusion-based acoustic models.
Although belonging to the ODE generative model family, flow matching with the rectified flow can automatically straighten its sampling trajectory, thus greatly lowering the sampling cost for generating high-quality speech.
Experiments on single and multi-speaker benchmarks proved the competence of VoiceFlow across different number of sampling steps, and the effectiveness of flow rectification for efficient generation.
We believe that the potential of flow matching in TTS research has yet to be discovered, including areas such as automatic alignment search and voice conversion.

\vspace{-0.1in}
\section{Acknowledgement}
\vspace{-0.05in}
This work was supported by China NSFC Project (No. 92370206), Shanghai Municipal Science and Technology Major Project \\(2021SHZDZX0102) and the Key Research and Development Program of Jiangsu Province, China (No. BE2022059).
\newpage
\bibliographystyle{customIEEEtran}

\bibliography{refs}

\end{document}